\documentstyle[11pt]{article}

\title{\bf Reggeized Gluons with a Running Coupling Constant}
\author{M. A. Braun  \\ Department of high-energy physics,
\\ University of S. Petersburg, 198904 S. Petersburg
, Russia and\\
Department of Particle Physics, University of Santiago de Compostela,\\
15706 Santiago de Compostela, Spain
} \date{July 1994}
\def\beq{\begin{equation}}
\def\eeq{\end{equation}}
\def\noi{\noindent}
\def\oq{\omega(q)}
\def\oa{\omega(q_{1})}
\def\ob{\omega(q_{2})}
\def\eq{\eta (q)}
\def\ea{\eta (q_{1})}
\def\eb{\eta (q_{2})}
\def\ec{\eta (q'_{1})}
\def\ed{\eta (q'_{2})}
\begin{document}
\maketitle
\medskip
\noi{\bf Abstract}
The equation for two reggeized gluons in the vacuum channel is
generalized to take into account the running QCD coupling  constant on
the basis of the bootstrap condition for gluon reggeization. Both the
gluon trajectory as a function of momentum and the interaction as a
function of distance grow like $\log\log$ in the ultraviolet. The
resulting equation depends on the confinement region. With a simple
parametrization of its influence by an effective gluon mass
the pomeron intercept turns out much smaller than for a fixed coupling
constant (the BFKL pomeron).

\vspace*{3 cm}
{\Large\bf US-FT/12-94}
\newpage
\section{Introduction}

Recently much effort has been spent generalizing the BFKL pomeron [ 1 ]
to include the running QCD coupling constant [ 2 ]. As in the first
attempts by Lipatov et al. [ 3 ] the idea is to introduce the running constant
directly into the equation for the vacuum channel on the basis of more
or less plausible assumptions about the momentum or distance scale at
which the gluons interact. Such an approach ignores the origin of the BFKL
equation, which is, in fact, a direct consequence of the assumption
that the gluon reggeizes. This assumption can be formalized in the
so-called bootstrap condition [ 3 ]: in the gluon channel the two-gluon
equation should possess a solution in the form of the gluon Regge pole.
One can then verify that with the BFKL interaction kernel and the known
gluon Regge-trajectory this condition is indeed satisfied.

The bootstrap condition is the basic assumption in deriving the BFKL
pomeron, which may be considered as its byproduct in relation to the
vacuum channel. For that reason, in our opinion, all attempts to change
the BFKL equation in order to introduce the running coupling constant
should start with the bootstrap equation rather than with the vacuum
channel one. One has to change simultaneously both the gluon interaction
and its trajectory in a manner consistent with the bootstrap
requirement. The present note is devoted to this problem.

We first prove that the bootstrap equation admits more general solutions
than the known BFKL one. These involve a generalized interaction kernel,
parametrized by a function of a single momentum variable, related to
the gluon Regge trajectory via a nonlinear integral equation. We
determine both the kernel and the trajectory asymptotically from the
known behaviour of the gluon density with the running coupling constant.
As a result the trajectory turns out rising like $\ln\ln q^{2}$ at high
momenta rather than like $\ln q^{2}$ as with a fixed coupling constant.
The behaviour of both the kernel and the trajectory at low momenta
$q^{2}\leq\Lambda^{2}$ remains undetermined and governed by the
confinement. However the bootstrap equation guarantees that all the good
properties of the interaction kernel for reggeized gluons remain the
same as for the fixed coupling constant BFKL case: the kernel results
insensitive to the infrared region. This leaves some hope that the
results will depend only weakly on the details of the confinement
mechanism.

Parametrizing the confinement effects by a single parameter (the "gluon
mass" $m$) and interpolating both the trajectory and the kernel between
the ultraviolet and infrared regions by their
asymptotic expressions at high momenta appropriately modified to match
the low momentum behaviour
 we investigate the resulting
equation for the pomeron at $q=0$. Contrary to the fixed constant case
the pomeron singularity now becomes a pole in the $j$ plane whose
position depends on the single dimensionless parameter $m/\Lambda\geq
1$. In the limit $m\rightarrow\Lambda$ the equation becomes singular and
effectively coincides with the BFKL one.  With growing $m$, as
variational calculations indicate, the pole moves to the left
 so that the intercept lies much closer to unity than for the BFKL
pomeron. According to these variational estimates,  the pomeron
intercept (minus one) $\Delta$ with the running coupling constant
for reasonable values of $m/\Lambda$ weakly depends on this ratio
and lies in the region of 0.15 -- 0.17.

\section{The bootstrap equation and its solution}

The bootstrap condition is a result of the requirement that the equation
for two reggeized gluons have a Regge pole solution in the gluon
channel. Let $\alpha(q)=1+\oq$ be the gluon Regge trajectory and
$K(q,q_{1},q'_{1})$ be the two-gluon interaction kernel. The
(homogeneous) equation for two reggeized gluons in colour representation
$R$ is then
\beq
(j-1-\oa-\ob)\psi(q_{1})=
\lambda_{R}\int(d^{2}q'_{1}/(2\pi)^{2})
K(q,q_{1},q'_{1})\psi(q'_{1})
\eeq
where $q=q_{1}+q_{2}$ is the total momentum, $j$ is the angular
momentum and $\lambda_{R}$ is the strength of the interaction for a given
colour representation. For colour group $SU(3)$ in the gluon channel
$\lambda_{8}=3/2$ and in the vacuum channel
$\lambda_{1}=2\lambda_{8}=3$. Concentrating on the gluon channel and
requiring that the solution of (1) be of the form of a Regge pole with
a trajectory $1+\oq$, one arrives at the bootstrap equation
\beq
(\oq-\oa-\ob)\psi(q_{1})=
\lambda_{8}\int(d^{2}q'_{1}/(2\pi)^{2})
K(q,q_{1},q'_{1})\psi(q'_{1})
\eeq
It is a complicated functional-integral equation which relates $\omega$,
$K$ and $\psi$. Following the known result for the BFKL pomeron we
shall seek solutions of (2) with a constant $\psi(q_{1})$. Then the
bootstrap equation simplifies to a relation between $\omega$ and $K$:
\beq
\oq-\oa-\ob=
\lambda_{8}\int(d^{2}q'_{1}/(2\pi)^{2})
K(q,q_{1},q'_{1})
\eeq

It is easy to find that this relation can be satisfied provided $K$ is
chosen in the form
\beq
K(q,q_{1},q'_{1})=(\ea/\ec+\eb/\ed)/\eta(q_{1}-q'_{1})-\eq/\ec\ed
\eeq
where the function $\eq$ is related to the trajectory $\oq$ by a
nonlinear integral equation
\beq
\oq=-\lambda_{8}\eq\int (d^{2}q_{1}/(2\pi)^{2})/\ea\eb
\eeq
Equations (4) and (5) allow to find the kernel $K$ for a given gluon
trajectory $\oq$ or vice versa. The BFKL pomeron results from a
particular solution of (4) and (5) with
\beq
\eq=2\pi q^{2}/g^{2}
\eeq
where $g$ is the fixed coupling constant.

 The form (4) garantees that if $\eq\rightarrow 0$
as $q\rightarrow 0$ the kernel $K$ remains nonsingular at
$q'_{1}=q'_{2}=0$ and goes to zero as $q_{1},q_{2}\rightarrow 0$.
One can observe that the leading contributions of the two terms in (4)
cancel in these limits. One can also prove that, with (4) and (5)
satisfied, the equation for two gluons in the vacuum channel
is nonsingular at $q_{1}-q'_{1}\rightarrow 0$. To do that use the
trick of [ 3 ] and represent the trajectory (5) as
\beq
\oq=-2\lambda_{8}\eq\int (d^{2}q_{1}/(2\pi)^{2})/\ea(\ea+\eb)
\eeq
Then, remembering that $\lambda_{1}=2\lambda_{8}$, in the vacuum
channel Eq. (1) can be written as
\[
(j-1)\psi(q_{1})=\lambda_{1}\int (d^{2}q'_{1}/(2\pi)^{2})
((\ea/\eta (q_{1}-q'_{1}))(\psi(q'_{1})/\ec-\psi(q_{1})/(\ec+\eta(q_{1}-
q'_{1}))\]
\beq
+(1\leftrightarrow 2)-\eq\psi(q'_{1})/\ec\ed)
\eeq
With $\eta(0)=0$ the singularity at  $q_{1}=q'_{1}$ on the right-hand
side is evidently cancelled between the terms coming from the kernel $K$
and the trajectory $\omega$. This cancellation has been long known for
the BFKL pomeron.

\section{The running coupling constant}

According to the considerations of the preceding section to generalize
for the running coupling constant one has to find with it either the
Regge trajectory $\oq$ and then construct the kernel $K$ using (4) and
(5) or directly the kernel $K$ with the structure (4), that is, the
function $\eq$, and then find the trajectory from (5). We shall use the
latter approach. To determine $\eq$ we consider the vacuum channel
equation (8) at $q=0$, which is equivalent to studying the gluon
distribution for virtuality $q_{1}^{2}=q_{2}^{2}$:
\[
(j-1)\psi(q_{1})=
2\lambda_{1}\int (d^{2}q'_{1}/(2\pi)^{2})
((\ea/\eta (q_{1}-q'_{1}))(\psi(q'_{1})/\ec \]
\beq
-\psi(q_{1})/(\ec+\eta(q_{1}-
q'_{1}))-\eq\psi(q'_{1})/\ec^{2})
\eeq
Let $q_{1}\rightarrow\infty$. Then the dominant (logarithmic)
contribution comes only from the first term, from the integration region
$q'_{1}<q_{1}$. In this limit the equation becomes
\beq
(j-1)\psi(q_{1})=2\lambda_{1}\int (d^{2}q'_{1}/(2\pi)^{2})
\theta(q_{1}^{2}-{q'}_{1}^{2})\psi(q'_{1})/\ec
\eeq
Comparing this with the well-known behavior of the gluon distribution at
large $q_{1}^{2}$ we find asymptotically
\beq
1/\eq=g^{2}(q)/2\pi q^{2}=2\pi/bq^{2}\ln q^{2}/\Lambda^{2}
\eeq
where $g(q)$ is the QCD running coupling constant,
\beq
b=(11-(2/3)N_{F})/4
\eeq
and $\Lambda\simeq 200\ Mev$ is the standard QCD parameter. Relation
(11) is valid for $q^{2}>>\Lambda^{2}$. So the comparison to the usual
evolution equation allows to determine $\eq$ for large values of the
argument.

The equation for two reggeized gluons (1) involves  all values of
$q^{2}$, smaller than $\Lambda^{2}$ included. As we have observed in
Sec. 2, the equation is stable respective to possible singularities of
$1/\eq$ as $q\rightarrow 0$. However it does not mean that the equation
does not depend on small values of $q^{2}$ at all. At most we may expect
that this dependence is weak. Thus we are faced with the problem of
prolongation of the asymptotical expression for $\eq$ (and the kernel
$K$) into the region of $q^{2}$ comparable to $\Lambda^{2}$ and even
smaller than $\Lambda^{2}$. This cannot be done on the basis of any
theoretical considerations. The region of small $q^{2}$ is that of
confinement where, in fact, the equation for two gluons itself looses
sense. Rather we have to invoke some initial values for the function
$\eq$ at small $q^{2}$ which effectively take into account the influence
of confinement.

The asymptotical expression (11) blows up at $q^{2}=\Lambda^{2}$ (the
infrared pole). A simple way to avoid it is to introduce "the gluon
mass" $m$ changing $q^{2}$ to $q^{2}+m^{2}$. Thus we assume for all
$q^{2}$
\beq
1/\eq=2\pi/b(q^{2}+m^{2})\ln ((q^{2}+m^{2})/\Lambda^{2})
\eeq
In this formula $\Lambda$ is the observable QCD parameter and $m$ is a
new parameter which incorporates the confinement effects. Evidently to
avoid the infrared pole we have to choose $m>\Lambda$. The limiting case
is when $m=\Lambda$ and the pole appears at $q=0$. As mentioned due to
the properties of the kernel the vacuum channel equation remains valid
for this especially simple case with no more parameters than $\Lambda$.
As calculation show, values $q^{2}<<\Lambda^{2}$ are essential in the
vacuum equation with $m=\Lambda$. Inspection of (13) then shows that
this limit is equivalent to the BFKL pomeron with a coupling constant
$g^{2}=2\pi/b$. One should keep in mind, however, than there are
no {\it a priori} theoretical arguments that might support the choice
$m=\Lambda$ and that the values $q^{2}<<\Lambda^{2}$ essential in this
limit are completely unphysical (deep inside the confinement region).
Thus the BFKL pomeron appears as a formal but unphysical limit of the
equation with a running coupling constant with $\ea$ given by (13).

\section{The pomeron at $q=0$}

At $q=0$ the equation (1) can be rewritten in a simpler form changing
the function $\psi$ according to
\beq
\psi(q)\rightarrow \eq\psi(q)
\eeq
Introduce the two-gluon potential $V$ with a kernel in the momentum space
\beq
V(q)=1/\eq
\eeq
Then we obtain an equation
\beq
(1-2\eq)\psi(q)=2\lambda_{1}\int
(d^{2}q'/(2\pi)^{2})V(q-q')\psi(q')-\lambda_{1}\eta(0)V(q)
\int(d^{2}q'/(2\pi)^{2})V(q')\psi(q')
\eeq
It has the form of the stationary Schroedinger equation in two
dimensions
\beq
H\psi=E\psi
\eeq
where the energy is $E=1-j$ and the Hamiltonian is
\beq
H=-2\oq-2\lambda_{1}V+\lambda_{1}\eta(0)|V><V|
\eeq
 The rightmost
singularity in $j$ corresponds to the ground state energy of the
Hamiltonian (18).

In the Hamiltonian (18) the Regge trajectory $\oq$ plays the role of the
kinetic energy. The interaction consists of two parts. The first one is
a normal local pair interaction with a potential
\beq
V(r)=\int (d^{2}q/(2\pi)^{2})V(q)\exp iqr
\eeq
The second part
(the last term in (18)) is a separable interaction, which
appears in our case because
according to (13) $\eta(0)\neq 0$ unless $m=\Lambda$. With $\eq$ given
by (13) both $\oq$ and $V$ are completely determined by (5) and (15).
However their explicit expressions in terms of momentum space integrals
are rather complicated and can be evaluated only by numerical
calculations. To have some idea about the behaviour of $\oq$ and $V(r)$
we therefore start by studying their asymptotical expressions for small
and large values of their corresponding arguments.

In the following we assume $\Lambda=1$.

Let us begin with the Regge trajectory. Its explicit expression is
given by (5) and (13):
\beq
\oq=-(\lambda_{8}/2\pi b)(q^{2}+m^{2})\ln(q^{2}+m^{2})\int d^{2}q_{1}/
(q_{1}^{2}+m^{2})\ln(q_{1}^{2}+m^{2})(q_{2}^{2}+m^{2})\ln(q_{2}^{2}+m^{2})
\eeq
with $q_{1}+q_{2}=q$. Evidently $\oq<0$ for all $q$. At $q=0$ it is
finite:
\beq
\omega(0)=-(\lambda_{8}/2b)(1+m^{2}\ln m^{2}{\mbox Ei}(-\ln m^{2}))
\eeq
Note that at $m=1$ the second term in (21) is zero and $\omega(0)=
-\lambda_{8}/2b$.
The asymptotical behaviour at $q^{2}>>1$ is provided by the
integration over $q_{1}^{2}<<q_{2}$ or $q_{2}^{2}<<q^{2}$ in (20). To
study it the form (7) is more convenient. Simple calculations give the
asymptotical expression
\beq
\oq_{q^{2}>>1}\simeq-(\lambda_{8}/b)\ln(\ln q^{2}/\ln m^{2})
\eeq
where we have retained the $\ln\ln m^{2}$ term to show the divergence at
$m=1$. As in the BFKL case this divergence is cancelled by the
contribution from the interaction in Eq. (1).

The potential $V(r)$ is explicitly given by
\beq
V(r)=(1/2\pi b)\int d^{2}q\exp iqr/(q^{2}+m^{2})\ln(q^{2}+m^{2})
\eeq
It goes to zero as $r\rightarrow\infty$
\beq
V(r=\infty)=0
\eeq
The asymptotical behaviour at small $r$ is determined by the behavior of
$\eq$ at large $q^{2}$. Calculations give
\beq
V(r)_{r<<1}\simeq (1/2b)\ln(\ln(1/r^{2})/\ln m^{2})
\eeq
As mentioned, in Eq.(1) the terms proportional to $\ln\ln m^{2}$ from (22)
and (25) cancel due to the relation $\lambda_{1}=2\lambda_{8}$.

As one observes, the Hamiltonian (18) possesses quite normal features
unless $m=1$. The kinetic energy $-2\oq$ results positive and steadily
grows with $q^{2}$ from its value $E_{0}$ at $q=0$ equal to
$-2\omega(0)$ and given by (21), which only leads to a constant energy
shift. The potential energy $-2\lambda_{1}V(r)$ is negative which means
attraction between the gluons, partially compensated by the repulsive
separable term. All interaction goes to  zero at large values of
$r$. At $r\rightarrow 0$ the potential goes to minus infinity, although
the singularity at $r=0$ is quite weak. With such an Hamiltonian we
expect normal bound states at negative values of the shifted energy
$\tilde{E}=E-E_{0}$ which means poles in the $j$-plane.

The obtained asymptotic expressions for $\oq$ and $V(r)$ are of course
valid only for large values of $q$ and $1/r$ respectively. The exact
form of both kinetic and interaction energy has to be calculated
numerically from (20) and (23) for a given $m$. We know, however, that on
the one side, the equation (1) is not very sensitive to infrared
behaviour of $\oq$ and $V(r)$ and, on the other hand, the initial
function $\eq$ can only be well determined at $q^{2}>>1$, so that only
the asymptotic expressions for $\oq$ at $q^{2}>>1$ and $V(r)$ at
$1/r^{2}>>1$ are well established. For these reasons, instead of using
the exact values of $\oq$ and $V(r)$  folllowing from
(20) and (23), we shall use their asymptotical expressions (22) and (25)
for all $q$ and $r$
respectively, appropriately modifying them to take into account the known
values of $\omega(0)$ and $V(r=\infty)$. Namely we shall use Eq. (17) with
$E$ changed to $E-E_{0}$ and with the expressions
\beq
\oq =-(\lambda_{8}/b)\ln(\ln(q^{2}+m^{2})/\ln m^{2})
\eeq
\beq
V(r)=(1/2b)\ln(\ln(1/r^{2}+m^{2})/\ln m^{2})
\eeq
which agree with (22) and (25) and go to zero at $q=0$ and
$1/r=0$, respectively. The constant term $\ln\ln m^{2}$ is in fact
cancelled between the first and second term in (18). In the limit
$m=1$ the last (separable) term in (18) disappears and the Hamiltonian
takes a simple form
\beq
H=(2\lambda_{8}/b)(\ln\ln(1+q^{2})-\ln\ln(1+1/r^{2}))
\eeq
As mentioned, in this limit small values of $q$ and $1/r$ turn out to
be essential. Then the Hamiltonian further simplifies to
\beq
H=(2\lambda_{8}/b)(\ln q^{2}+\ln r^{2})
\eeq
which is nothing but the BFKL Hamiltonian  up to  an energy shift.

\section{Variational calculations of the pomeron intercept}

For all its nice look, Eq. (17) with kinetic and interaction terms given
by (26) and (27) is much more difficult to solve than the BFKL one due to
the lack of scale invariance. So instead of trying to solve it exactly
we investigate the ground state energy by the standard variational
technique. Evidently the lowest energy eigenvalue $E=1-j$ depends only
on the dimensionless ratio of the gluon mass $m$ to the QCD parameter
$\Lambda=1$.

We have taken our trial functions as linear combinations of the
two-dimensional harmonic oscillator functions for zero angular momentum
\beq
\psi(r^{2})=\sum_{k=0}^{N} c_{k}{\mbox L}_{k}(ar^{2})\exp (-ar^{2}/2)
\eeq
with ${\mbox L}_{k}$ the Laguerre polinomials and $a$ and $c_{k}$
variational parameters. Up to 10  polinomials have been included
in the present calculations.

The resulting ground state energies in units $\lambda_{1}/b$
 for different values of $m$ are shown in Table 1 (the second column).
 Considered
as a function of $m$ the ground state energy $E(m)$ has a sharp minimum
at $m=1$, which can be determined exactly from the known intercept of
the BFKL pomeron. With the growth of $m$ the energy first goes down
very steeply at small values of $m-1$ but then flattens and beginning
from $m-1\sim 0.1$ diminishes very slowly with $m$. For $m$ greater than
2 the obtained values for the ground state energy do not practically
depend on $m$ up to $m=6$, changing from 1.3 to 1.1.
Also the optimal values of the parameter $a$ are shown in Table 1
(the third column). These
characterize the typical values of $q^{2}$ for the gluons. For $m=1$
$a$ goes to zero: as mentioned small values of $q^{2}$ are essential
in this case, which makes it completely unphysical. However with the
growth of $m$ the average values of $q^{2}$ rapidly grow to become
considerably greater than 1 for $m^{2}$ greater than 1.1. Then the
picture
becomes self-consistent, restricted to the region of high momenta.
In the last column of Table 1
 the values of the intercept
are shown  for QCD with three flavours ($\lambda_{1}/b=4/3$).
For all values of $m$, except the ones quite close to 1, the intercept
results considerably smaller than for the BFKL pomeron. Its typical
values are
of the order of 0.15. It is remarkarble that  such values are close to
the ones usually taken for the pomeron intercept in the
 semiphenomenological description
of soft strong phenomena in the framework of the supercritical pomeron
model [ 4 ].

\section{Acknowledgments}

The author is deeply grateful to E. M. Levin, L. N. Lipatov,
C. Pajares and N. Armesto for
fruitful and constructive discussions. He also expresses his gratitude
to the General Direction of the Scientific and Technical Investigation
(DGICYT) of Spain for financial support.

\newpage

\section{References}
\noi 1. E. A. Kuraev, L. N. Lipatov and V.S. Fadin, Sov. Phys. JETP {\bf 45}
(1977) 199;\\
Ya. Ya. Balitzky and L. N. Lipatov, Sov. J. Nucl. Phys. {\bf 28} (1978) 822.\\
\noi 2. E. M. Levin and M. G. Ryskin, Phys. Rep. {\bf 189} (1990) 267;\\
N. N. Nikolaev and B. G. Zakharov, Phys. Lett. {\bf B327} (1994) 157;\\
J. Kwiecinski, A. D. Martin, P. J. Sutton and K. Golec-Biernat,
preprint DTP/94/08.\\
\noi 3. L. N. Lipatov in "Perturbative Quantum Chromodynamics", Ed.  A. H.
Mueller, World Scientific 1989.\\
\noi 4. A. Capella, U. P. Sukhatme, C.-I. Tan and J. Tran Thanh Van, Phys. Rep.
{\bf 236} (1994) 225;\\
A. B. Kaidalov and K. A. Ter-Martirosyan, Phys. Lett. {\bf B117} (1982) 247.
\newpage
\vspace*{3 cm}
{\large\bf Table}
\vspace {1 cm}

\begin{tabular}{|r|r|r|r|}\hline
$m^{2}/\Lambda^{2}$&$ E$& $a=<q^{2}>/\Lambda^{2}$& $\Delta$\\\hline
1.0& -2.04& 0.& 2.72\\\hline
1.01&-0.518& 0.50& 0.690\\\hline
1.05&-0.309& 1.50& 0.412\\\hline
1.1& -0.238& 2.56& 0.317\\\hline
1.5& -0.145& 11.3& 0.193\\\hline
2.0& -0.129& 24.0& 0.172\\\hline
3.0& -0.119& 54.4& 0.159\\\hline
4.0& -0.114& 90.0& 0.152\\\hline
6.0& -0.109&172.0& 0.145\\\hline
\end{tabular}
\vspace{1 cm}

{\Large\bf Table captions}
\vspace{1 cm}

The first column gives values of the gluon mass squared in units
$\Lambda^{2}$. In the second column ground state pomeron energies are
presented in units $\lambda_{1}/b$. The third column gives the
optimal values of the variational parameter $a$ which correspond to
average gluon momentum squared in units $\Lambda^{2}$. The last column
gives the pomeron intercept (minus one) for the QCD with three flavours
($b=9/4$).

 \end{document}